# CLARAPRINT: A CHORD AND MELODY BASED FINGERPRINT FOR WESTERN CLASSICAL MUSIC COVER DETECTION

**Author**
Mickaël Arcos
**mickael.arcos@rondodb.com**

## ABSTRACT

Cover song detection has been an active field in the Music Information Retrieval (MIR) community during the past decades. Most of the research community focused in solving it for a wide range of music genres with diverse characteristics. Western classical music, a genre heavily based on the recording of "cover songs", or musical works, represents a large heritage, offering immediate application for an efficient fingerprint algorithm. We propose an engineering approach for retrieving a cover song from a reference database thanks to a fingerprint designed for classical musical works. We open a new data set to encourage the scientific community to use it for further researches regarding this genre.

## 1. INTRODUCTION

In western classical music, it is of regular practice to play and record the same works of art overtime. Regarding works independently from recordings is of central interest for a wide range of applications. We propose a fingerprint mechanism and a series of experiments considering the original material not as another recording but as a western classical musical work. We leverage high-level features, harmony and melody, that we think are enough to capture the unicity of such a work.

Indeed, a classical musical cover (or *interpretation*) is more specific than a cover song in popular music, in particular:

- each cover records a work materialized by one written score. Differences between covers are much limited than the ones in popular music, and mostly include tempo and local artistic expressions such as rubato;
- there is no original or source recording. The origin of an interpretation is a written document and does not belong to the audio signal realm;
- features such as instrumentation, chord progression, melody progression, lyrics and tonality are extremely similar, if not identical, from one cover to another;
- descriptive metadata, such as title and composer name, can vary a lot because of translations, transliterations and partial titles from one cover to another;
- the complexity of melodies, harmonies and instrumentation makes it harder to tackle with generic approaches;
- many works do not follow an explicit beat pattern, limiting beat-tracking based feature extraction efficiency.

For the rest of this paper we will use the *classical music* expression for *western classical music*, excluding classical art music from other regions of the world that may not share the same characteristics.

This paper is structured as follow: section 2 gives an overview of related research work on audio fingerprinting and cover detection; section 3 describes the new data set released with this research; section 4 details the engineering proposition to build, store and retrieve the *claraprint*, as used in the experiments commented in section 5. We conclude in section 6 while suggesting tracks for future related work.

## 2. RELATED WORK

Audio fingerprinting has been an active research topic since the amount of digital audio recordings kept growing. The need for an efficient way to retrieve the exact recording or similar ones became more and more urgent. The range of applications is wide, including file searching, automated radio monitoring for broadcasting, copyright infringement verification, duplicate file detection, cover detection and automated indexing of large-scale audio catalogues. These applications led to a various propositions for audio fingerprinting [15].

Fingerprinting methods aim at looking for a recording in a database given an unknown recording, in the same fashion as human fingerprints. Low level features are often used to capture perceptual audio similarities from the audio signal, such as the Philips robust hash (PRH) used in [13], [14] and extended in [16]; Mel-Frequency Cepstrum Coefficients (MFCC) ([18], [20]); Spectral Flatness Measure (SFM) ([19]); Constant-Q-Transform (CQT) ([11]); spectral subband moments ([12]); bands with prominent tones ([22]) or with energy ([23], [7]); or harmonicity, bandwidth, loudness ([18]).

Some other systems include high-level musically meaningful attributes, like chroma ([6], [29]) and beat-synchronous chroma ([27]); rhythm ([17]); prominent pitch ([18]); chord recognition results, using Hidden Markov Model (HMM) ([8]) or rule-based model ([9]). Global redundancies within a song are used in [19], [20] and [21]. It is to be noted that [27] focuses in fingerprinting classical music recordings.

Cover song detection is another specific research field and includes techniques such as cross-correlation based identification system with dynamic beat tracking ([10]); metadata such as title and lyrics ([25]); approximate chord detection ([30], [31]); hashed-chroma landmarks ([32]); or 2D Fourier Transform Magnitude ([33]).

The research performed in [28] gives an interesting angle to a classical music melody detection. It was not integrated in this research as it mainly focuses on symphonic music (excluding vocal and instrumental music as present in our data set) and aims at a very accurate melody transcription which is not the heart of our research.

This related research focuses in either fingerprinting and retrieving audio *recordings* efficiently, either finding the best similarity algorithm in a pair-wise comparison, less prone to scalability; or do not focus or report on classical music specifically. In this research we are joining the two fields of fingerprinting and cover detection to match audio covers in an efficient and scalable way that captures the relevant musical information in respect to the particularities of classical music. We base our fingerprint on high-level features (chord and melody) computed with recent algorithms ([1], [2], [4], [24]). The proposed fingerprint system respects the main requirements as described in [15]: captures a perceptual digest of the recording, invariant to distortions, compact and easily computable.

## 3. DATA SET

We introduce a new classical music metadata data set containing a list of 100 works, with five interpretations for each via Youtube links[1]. This data set has the following characteristics:

- the list of works has been manually created and annotated;
- the selection is a diverse catalogue of famous and less famous classical music works[2];
- the five interpretations per work are all from different recordings;
- each recording contains a *live* (recorded live) and a *start_at* manual annotations (skipping applauses, instrument tuning or other artefacts);
- each work contains descriptive metadata such as opus numbers, detailed titles, composer names and dates, tonality and a link to Wikipedia.

Each work, or *clique,* is mostly recorded with the same instrumentation despite rare examples such as *Jean-Philippe Rameau: Suite en sol in G major "Nouvelle suite" - 6. Les sauvages* played either by a harpsichord or a piano; and the same tonality or very close ones such as in *Ludwig Van Beethoven: Symphony No.5 in C minor, Op. 67* in a range of about 1 tone away from C due to historical and artistic considerations.

This data set is provided along this research paper, in JAMS format [26].

## 4. METHODOLOGY

The current research is based on two complementary hypothesis:

1 https://zenodo.org/record/3911754#.X0oVlIvgpPZ

2 However it does not contain 12-tone works or fully unmelodic works such as percussive-only.

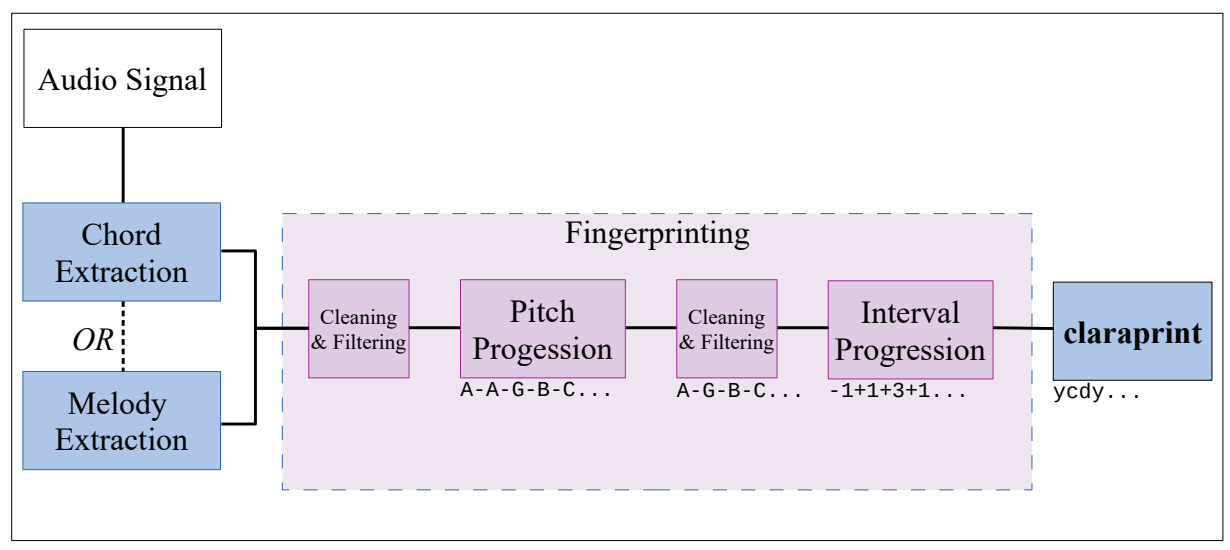

**Figure 1.** An overview of a claraprint generation.

1. given a sufficient amount of recorded time, the chord progression and the melody sequence of classical music works are discriminative enough to be used as the root of a fingerprint;
2. even if the chord extraction and melody extraction do not match accurately the audio signal, the inaccuracies do not prevent the similarity search from returning valid results.

In this section we describe the building of the fingerprint (4.1), the used algorithms to retrieve musical information (4.2), and the storage and retrieval of similar fingerprints (4.3).

### 4.1. Building the claraprint

Figure 1 shows the generation of a claraprint based on a chord or melody extraction algorithm:

1. the **audio signal** is truncated down to the first 30 or 120 seconds in the scope of this research. Applauses, silence, noise or any other non-musical introductory elements are skipped;
2. the **chord and melody extraction algorithms** are a third party contribution, and could be substituted or extended to other algorithms than the ones used in this research;
3. the first **cleaning & filtering** pass will adjust the previous output deleting undefined or low confident values and removing precise time information to only keep the order;
4. the **pitch progression** reduces the previous output to a simple progression. In the case of the chord extraction, each chord is simplified (*D* instead of *Dmaj7,* or *C* for *Cmin* or *Cmaj* for example); in the case of the melody extraction, the octave information is removed;
5. the second **cleaning & filtering** pass deletes repetition in the pitch progression;
6. the progression is converted to a succession of **intervals**. Each interval is the closest way to go from one note to the other, going up or down;
7. ultimately, each interval is converted into a **single letter** from a given dictionary, making the resulting string prone to near string matching techniques. The chord intervals dictionary (*[a-n]*) does not intersect with the melody one (*[o-z$%]*) which make them combinable.

Most of these steps will remove information from the original audio signal, only to keep the core of its melodic and chord progression information.

### 4.2. Used MIR algorithms

*Chords Extraction*

**“Chordino”** (abbreviated to **ch** in the rest of this paper) is a chord-extraction algorithm that uses a non-negative least squares (NNLS) based approximate note transcription, prior to a chroma mapping [1]. The resulting chroma feature is called NNLS chroma. The positive influence of the approximate transcription is particularly strong on chords whose harmonic structure causes ambiguities, and whose identification is therefore difficult in approaches without prior approximate transcription.

**“Crema” (cr)** is a chord recognition model based on the structured prediction model of McFee and Bello [2]. The used implementation has been enhanced to support inversion (bass) tracking, and predicts chords out of an effective vocabulary of 602 classes. Chord class names are based on an extended version of Harte’s [3].

*Melody Extraction*

**“Melodia” (me)** is based on the creation and characterisation of pitch contours. Time continuous sequences of pitch candidates are grouped using auditory streaming cues as described in [4]. It is defined by a set of contour characteristics and shows that by studying their distributions it is possible to devise rules to distinguish between melodic and non-melodic contours. This led to the development of new voicing detection, octave error minimisation and melody selection techniques.

**“Piptrack” (mp)** is a pitch tracking on thresholded parabolically-interpolated STFT algorithm[1]. The used implementation is the one available in the Librosa python music library[2][24].

### 4.3. String matching and similarity

As explained in section 4.1, a claraprint is a succession of letters, each representing an interval between two pitches. For the querying purpose, this long word is *w*-shingled, e.g. divided into words of length *w*. For example, given $w=4$, the claraprint *abfhqjfui* becomes $\{abfh, bfhq, fhqj, hqjf, qjfu, jfui\}$. Empirically, the best results were found with a multi-*w*-shingling where $w \in \{2,3,4,5,6,7\}$, written in the rest of this paper as *[2-7]-shingled* claraprint.

We used the Okapi BM25 similarity algorithm (BM is an abbreviation of best matching), a ranking function used by search engines to estimate the relevance of documents to a given search query. It is based on the probabilistic retrieval framework developed in the 1970s and 1980s by Stephen E. Robertson, Karen Spärck Jones, and others[3][5]. BM25 is a bag-of-words retrieval function that ranks a set of documents based on the query terms appearing in each document, regardless of their proximity within the document. It is a family of scoring functions with slightly different components and parameters. One of the most prominent instantiations of the

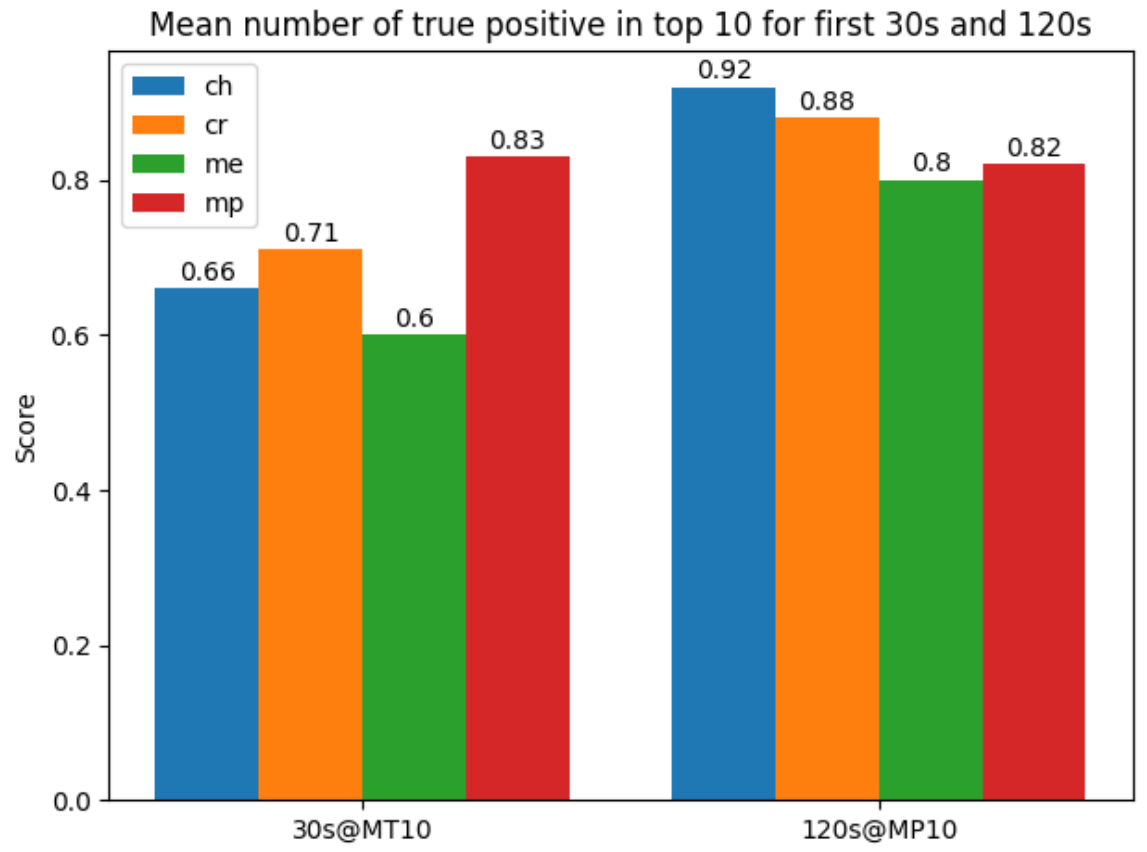

**Figure 2.** Mean value of true positive in top 10 for claraprints computed on the first 30s and 120s.

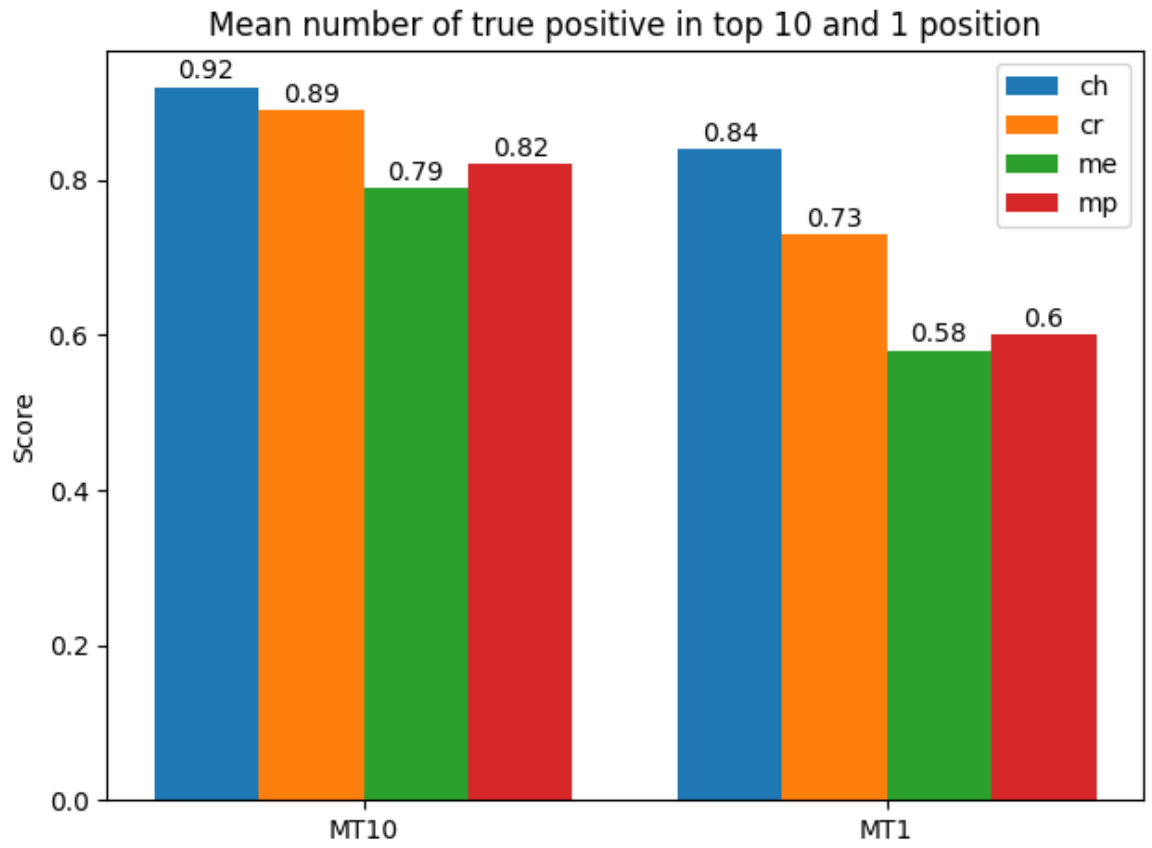

**Figure 3.** Mean value of true positive in top 10 (MT10) and first (MT1) position for 120s claraprints.

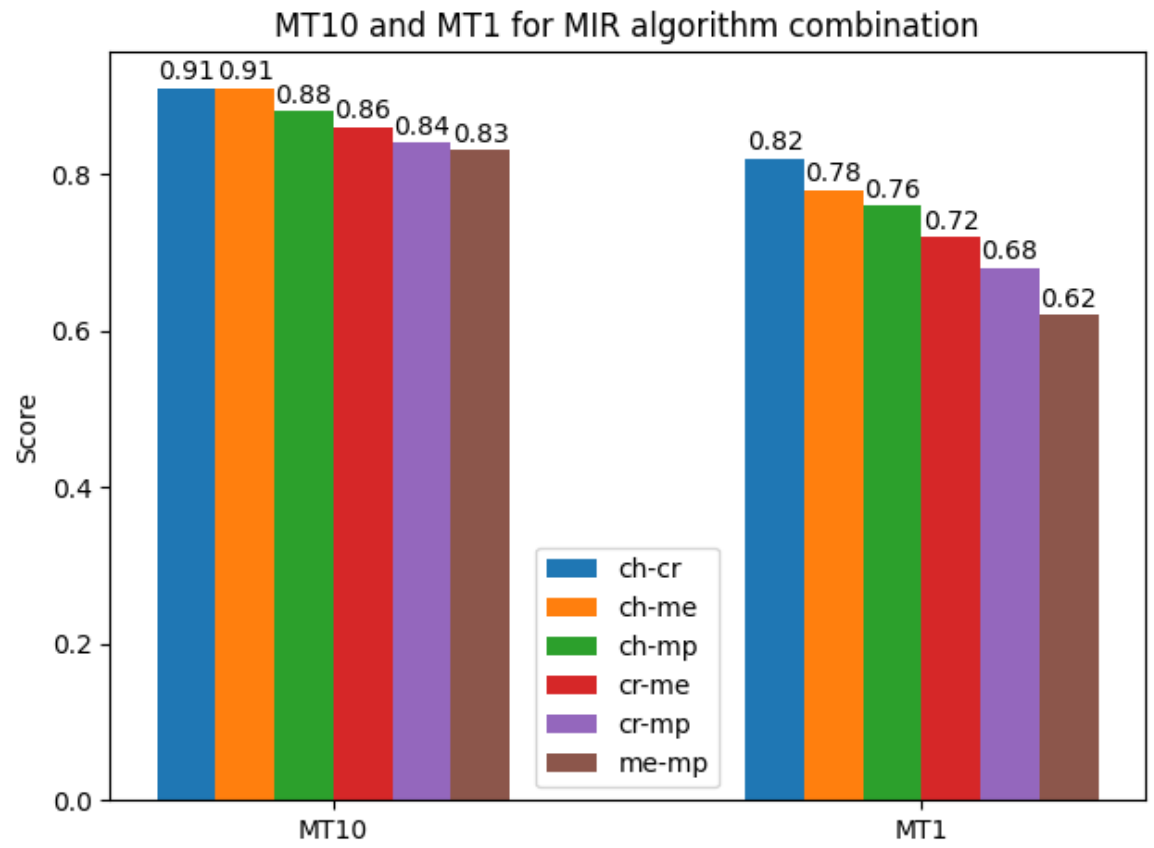

**Figure 4** Mean value of true positive in top 10 (MT10) and first (MT1) position of combined [2-7]-shingled claraprints.

1 https://librosa.github.io/librosa/generated/librosa.core.piptrack.html
2 https://librosa.github.io/librosa/
3 https://en.wikipedia.org/wiki/Okapi_BM25

function is as follows. Given a query Q, containing keywords $q_1,\ldots,q_n$, the BM25 score of a document $D$ is:

$$score(D,Q)=\sum_{i=1}^{n} IDF(q_i)\cdot\frac{f(q_i,D)\cdot(k_1+1)}{f(q_i,D)+k_1\cdot(1-b+b\cdot\frac{|D|}{avgdl})}$$

where $f(q_i,D)$ is $q_i$ 's term frequency in the document $D$, $|D|$ is the length of the document $D$ in words, and $avgdl$ is the average document length in the text collection from which documents are drawn. $k_1$ and $b$ are free parameters, usually chosen, in absence of an advanced optimization, as $k_1\in[1.2,2.0]$ and $b=0.75$. $IDF(q_i)$ is the inverse document frequency weight of the query term $q_i$. It is usually computed as:

$$IDF(q_i)=\log\frac{N-n(q_i)+0.5}{n(q_i)+0.5},$$

where $N$ is the total number of documents in the collection, and $n(qi)$ is the number of documents containing $q_i$. In our case, a *document* is a [2-7]-shingled claraprint and a *word* a short extract of a pitch progression. Each word will be lowered in the ranking result if it is often present in all our fingerprint corpus, e.g. if this progression pattern is often to be found in classical music works.

Elasticsearch[1] is a search engine based on the free and open-source Lucene[2] library. It provides a distributed, multitenant-capable full-text search engine with an HTTP web interface and schema-free JSON documents. It is the back end used for the experiments described in the section 5. We use Elasticsearch's Okapi BM25 algorithm implementation[3] with default $k_1=1.2$ and $b=0.75$.

## 5. RESULTS

In Figure 2 are displayed the mean value of true positive results in our corpus of fingerprints computed either on the first 30 seconds (left) or the first 120 seconds (right) of the audio recording. Globally the performances are greatly improved by the use of a longer time range. In classical music, it's common that the recordings are quite long compared to popular music. We find quite unsurprising that 30 seconds are not enough to capture the core of the fingerprint information. In particular, the chord progression can be very limited on such a duration, for slow paced or more static musical works. Using the full duration of the audio recording would require too much computation and generating a too long claraprint to keep it efficient. Some classical music work can last 30 minutes and even more.

Figure 3 shows that Chordino (ch) performs best at this task. For a randomly picked recording in each clique, all the other fingerprints are searched against the corpus. This step is repeated five times and averaged (which explains the negligible discrepancy between Figure 2 right and Figure 3 left). The chord-extraction based algorithms perform better than melody-extraction based algorithms. This could be explained by the bigger

1 https://www.elastic.co/elasticsearch/
2 https://lucene.apache.org/core/
3 https://www.elastic.co/guide/en/elasticsearch/reference/current/index-modules-similarity.html

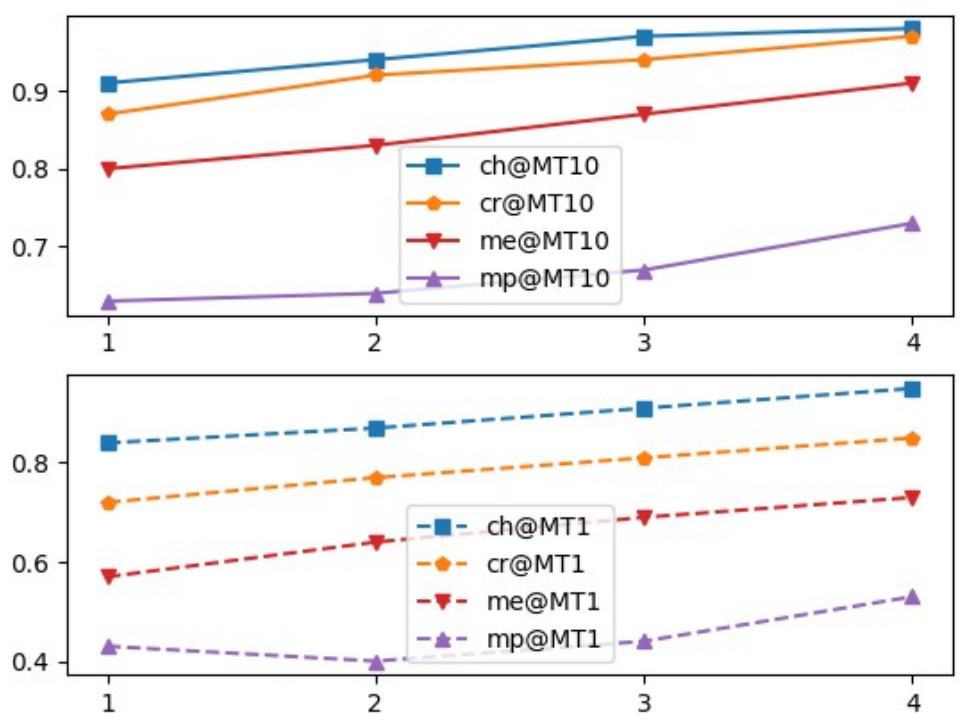

**Figure 5.** Mean value @10 and @1 of multiple-recording [2-7]-shingled claraprint combination, from 1 to 4 recordings of the same piece.

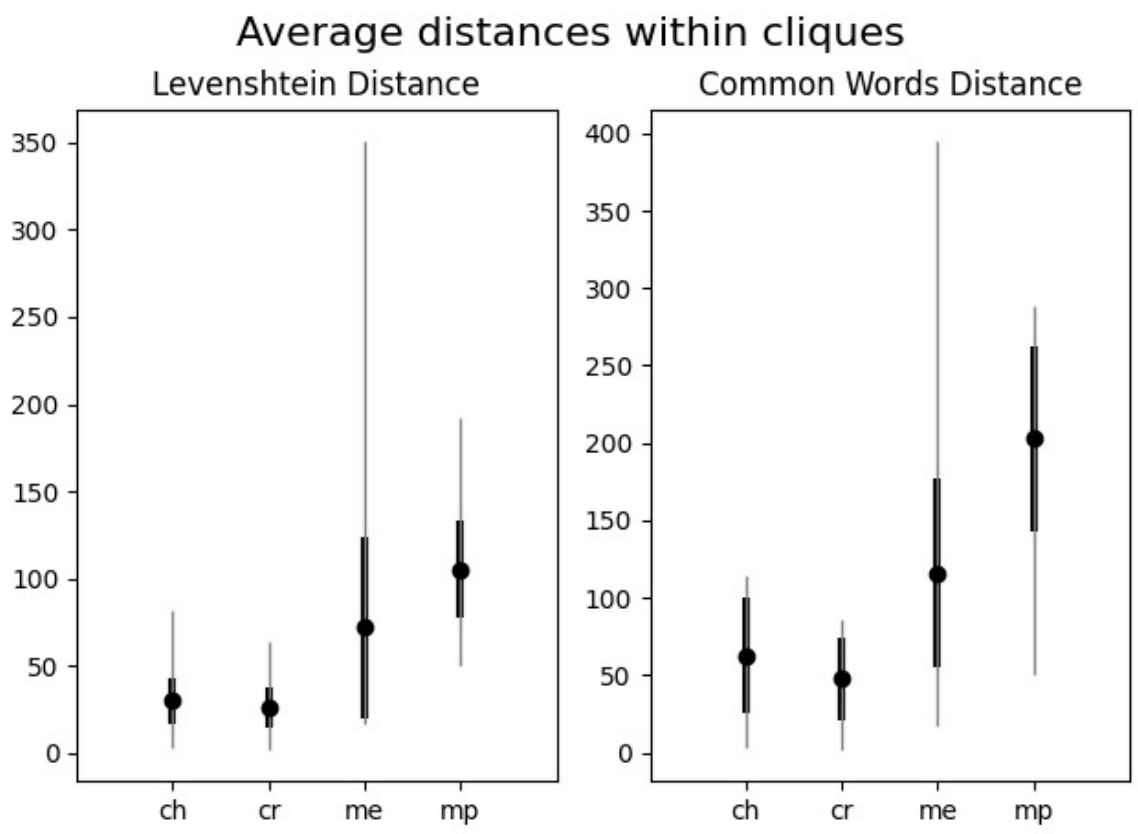

**Figure 6.** The average, minimal, maximal and standard deviation of pair-wise comparison of claraprints within the same clique, with the Levenshtein similarity algorithm (plain claraprints) and the common words in [2-7]-shingled claraprints.

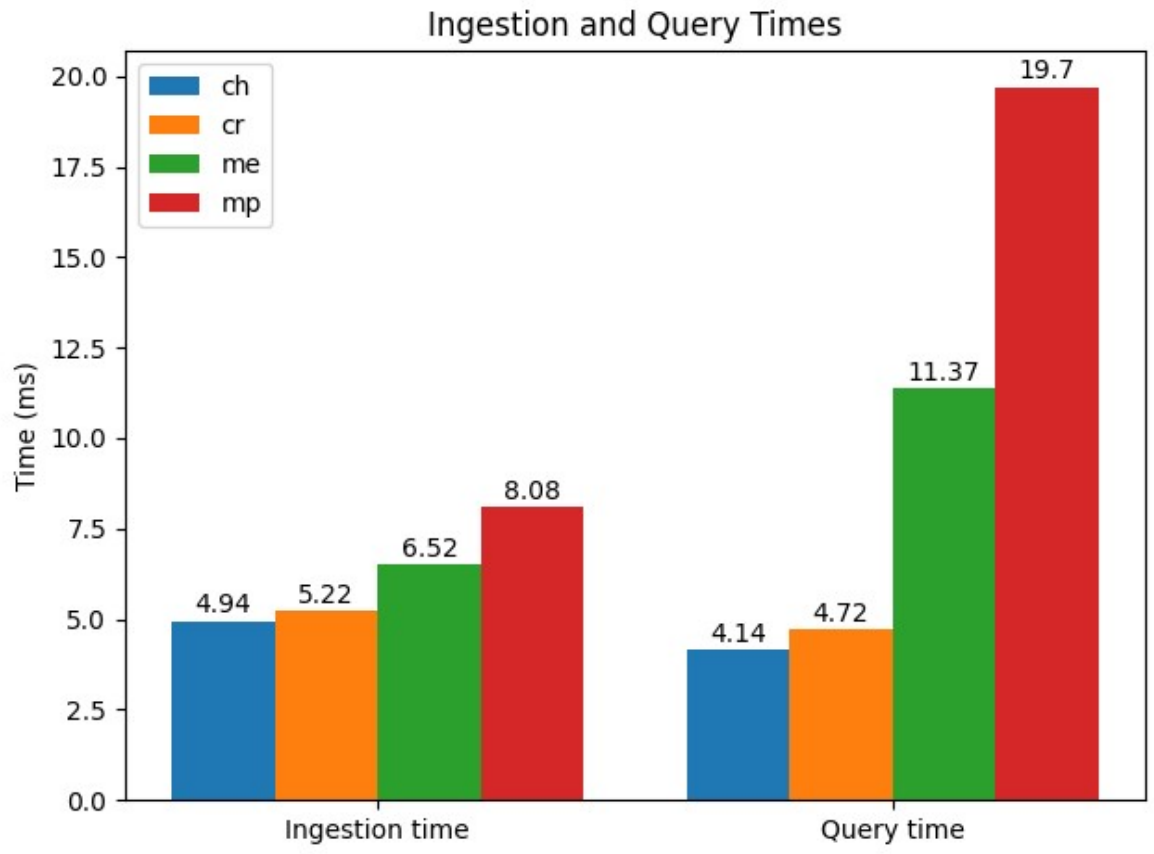

**Figure 7.** Ingestion and query times, in average, for each of the studied claraprint, each based on one recording.

average distance between two fingerprints within the same clique for melody-extraction based algorithms as shown in Figure 6.

In Figure 4, different algorithms were used to fingerprint the same work, and concatenated to generate a combined [2-7]-shingled claraprint. It is noticeable that the best results are not significantly improved by this combination. Chord-extraction based claraprints always perform better, while melody-extraction based claraprints search significantly improves when combined with a chord-extraction algorithm. The best combination is the two chord-extraction based claraprints, *ch* and *cr*.

In Figure 5, several recordings of the same clique were used to generate a combined [2-7]-shingled claraprint. The rest of the fingerprints in the clique were searched against the combined claraprints. Using several covers almost always improve the quality of the search results. This situation applies well in a context of an already partially annotated catalogue. Combining several claraprints of the same work forms a stronger reference claraprint.

Finally Figure 7 displays the average ingestion time and query time in milliseconds. Although all the values are very low, taking advantage of the Elasticsearch back end, we observe that melody based fingerprint are slower to ingest and query than the chord-based ones. This is explained by the difference of length between chord based claraprints versus melody based claraprints. It is indeed understandable that for a given recording the pitch sequence of a melody is longer than a chord progression sequence.

## 6. CONCLUSION AND FUTURE WORK

We presented a practical way of creating a fingerprint of recordings belonging to the same clique of western classical music works. We showed that using chord-extraction based algorithms performs better than melody-extraction based ones; and that combining various recordings of the same clique to generate a reference fingerprint noticeably improves these results.

For a future work we would like to extend the benchmark of the algorithms used to build claraprints. Using other features such as the main key, the duration of the recording, the instrumentation (annotated or extracted) can also be envisioned to make the fingerprint more discriminative. It could be interesting as well to automatically update an existing claraprint over time, given new matched recordings of a growing catalogue. Finally, extending the data set with 12-tone, atonal or percussive music would challenge the build of a robust claraprint for a comprehensive repertoire of classical music.

Along with these results, we release a new data set of a hundred works with five recordings each and precomputed data for four algorithms in the JAMS format. A framework written in python allows easy reproducibility of the presented results and could be extended to other pitch extraction algorithms[1]. We encourage the community to reuse this material to enhance the detection of classical music works and better address this genre in MIR research.

1 https://github.com/miqwit/claraprint